# High Temperature Ferromagnetism in GaAs-based Heterostructures with Mn Delta Doping

A. M. Nazmul,[1,2] T. Amemiya,[1] Y. Shuto,[1] S. Sugahara,[1] and M. Tanaka[1,2]

1. Department of Electronic Engineering, The University of Tokyo, 7-3-1 Hongo, Bunkyo-ku, Tokyo 113-8656, Japan.
2. PRESTO/SORST, Japan Science & Technology Agency, 4-1-8 Honcho, Kawaguchi, Saitama 332-0012, Japan.

Abstract:

We show that suitably-designed magnetic semiconductor heterostructures consisting of Mn delta ($\delta$)-doped GaAs and $p$-type AlGaAs layers, in which the locally high concentration of magnetic moments of Mn atoms are controllably overlapped with the 2-dimensional hole gas wavefunction, realized remarkably high ferromagnetic transition temperatures ($T_C$). Significant reduction of compensative Mn interstitials by varying the growth sequence of the structures followed by low temperature annealing led to high $T_C$ up to 250 K. The heterostructure with high $T_C$ exhibited peculiar anomalous Hall effect behavior, whose sign depends on temperature.

The introduction of bandgap engineering and wavefunction engineering into the study of collective magnetic phenomena in semiconductors is opening ways to manipulate the magnetic properties by changing the carrier characteristics such as the wavefunction, concentration, and carrier spin. For example, ferromagnetic order in the quasi 2-dimensional carrier systems of II-VI and III-V based magnetic semiconductor heterostructures can be induced or erased by changing the overlap of the 2-dimensional hole gas (2DHG) wavefunction with the magnetic impurities[1,2]. The magnetization and the ferromagnetic transition temperature ($T_C$) in these systems can be controlled by changing the carrier concentration with a gate-electric field or by changing the carrier-spins by circularly polarized light irradiation[3-5]. Despite such fundamental studies, the material design with $T_C$ near or above room temperature remained one of the most important challenges towards the realization of the semiconductor spintronics.

Theoretical calculations within the mean-field formalism predicted that the $T_C$ can be raised with increasing both the magnetic dopant concentration and the carrier concentration in semiconductors[6,7]. Experimental investigations were carried out to maximize the concentration of Mn substitutionally incorporated in the cation site of GaAs to act as acceptors rather than as compensating interstitial defects. It was reported that long-time annealing of (GaMn)As at low-temperature (LT) ($\leq 200°C$) can reduce the interstitial defect density, leading to significant improvement in its $T_C$ as high as 159 K[8]. It was also pointed out that the interstitial defect density can be controlled by changing its formation energy with changing the Fermi energy during growth[9]. In this letter, we show that suitably-designed $p$-type selectively doped III-V heterostructures ($p$-SDHS) with Mn delta ($\delta$)-doping in the 2DHG channel have ferromagnetic order with remarkably high $T_C$. Here, the $\delta$-doping of Mn atoms in semiconductors allows locally high concentration of magnetic moments exceeding the solubility limit in the bulk semiconductor host[10,11]. We found that the growth sequence of the *p*-SDHS with Mn $\delta$-doping and LT-annealing for suppressing interstitial defect formation are very important to realize high $T_C$. We also observed a clear sign-change in anomalous Hall effect (AHE) of the high $T_C$ heterostructure.

We have grown two types of *p*-SDHS samples with Mn $\delta$-doping (sample-A and sample-B) by molecular beam epitaxy (MBE) on semi-insulating (SI) GaAs(001) substrates. In sample-A, a GaAs layer with Mn $\delta$-doped content $\theta_{Mn}$ = 0.5 monolayer (ML) ($\theta_{Mn}$ = 1 ML corresponds to a sheet Mn concentration of $6.3 \times 10^{14}$ cm$^{-2}$ [11]) was grown at a substrate temperature $T_s$ of 300°C on a *p*-type Be-doped Al$_{0.5}$Ga$_{0.5}$As layer (Be concentration = $1.8 \times 10^{18}$ cm$^{-3}$, $T_s$ = 600°C), as shown in Fig. 1(a). Here, holes are supplied from the underlying *p*-AlGaAs layer to the overgrown Mn $\delta$-doped GaAs channel, resembling an inverted high electron-mobility transistor (I-HEMT). Details of the growth procedures are described elsewhere[11]. The thickness $d_s$ of the undoped-GaAs separation layer, which was a measure to control the interaction between the Mn $\delta$-doped GaAs layer and the 2DHG formed at the GaAs/*p*-AlGaAs interface, was set to 2 nm to enhance the overlap of the 2DHG wavefunction and the Mn $\delta$-doped layer[2]. Post-growth annealing was carried out at 300°C for 15 minutes in a N$_2$ atmosphere, which was optimum to maximize $T_C$ in our previous report[2]. On the other hand, in sample-B, a Mn $\delta$-doped ($\theta_{Mn}$ = 0.6 ML) GaAs layer was grown at $T_s$ = 285°C below a *p*-type Be-doped Al$_{0.3}$Ga$_{0.7}$As layer (Be concentration = $1.0 \times 10^{19}$ cm$^{-3}$), as shown in Fig. 1(b). Here, holes are supplied from the overgrown $p$-AlGaAs layer to the Mn δ-doped GaAs layer ($d_s$ = 0 nm), resembling a normal (N)-HEMT. Post-growth LT-annealing was carried out at 200°C for 112 hours in a $N_2$ atmosphere.

The structural properties of the Mn δ-doped heterostructures were characterized by high-resolution transmission electron microscopy (HRTEM). Figure 2(a) shows a cross-sectional HRTEM lattice image of a Mn δ-doped GaAs with $\theta_{Mn}$ = 1 ML prepared under the same conditions of sample-A, in which one can see no dislocation, and a slightly dark area with a width of ~2 ML is most likely due to strain induced contrast at the Mn δ-doped sheet. Also, Fig. 2(b) and (c) show cross-sectional HRTEM lattice images of the structure of sample-B, indicating that the $Al_{0.3}Ga_{0.7}As$/ GaAs interface (we see week contrast in (b)) is atomically abrupt, and one can see no clusters or dislocations in the whole structure. These studies suggest that pseudomorphic growth of Mn δ-doped sheets occurred maintaining the zinc-blende (ZB) type crystal structure, which is consistent with the reflection high energy electron diffraction patterns observed during the MBE growth of the samples.

The anomalous Hall effect (AHE) effect [12] has been used to study the magnetic properties of the present quasi-two dimensional $p$-SDHS systems for which bulk magnetization measurements are difficult. Hall measurements were carried out on patterned Hall bars with a channel width and a length of 50 μm and 200 μm, respectively. Au wire leads were soldered to the sample with In for ohmic contacts. The post-growth LT-annealing and annealing for ohmic contacts were done at 200°C or below, thus excluding the possibility of ferromagnetic MnAs second phase formation. Figure 3(a-e) show the Hall resistance $R_H$ as a function of magnetic field of sample-A. Here, in this system, $R_H = R_{O-sheet}B + R_{S-sheet}M$, where $R_{O-sheet}$ and $R_{S-sheet}(=cR_{sheet})$ are the anomalous and ordinary Hall coefficients, $B$ is the applied magnetic field, $M$ is the magnetization of the sample, $c$ is a constant, and $R_{sheet}$ is the sheet resistance. The second term is dominant in magnetic materials, thus $R_H \approx R_{S-sheet}M = cR_{sheet}M$. Clear ferromagnetic hysteresis at 130, 180, 185, and 190 K changed to a linear character at 205 K in Fig. 3(a)-(e), suggesting a phase transition from ferromagnetic to paramagnetic above 190 K. The positive linear component superimposed on the hysteresis could be due to a possible parallel conduction either in the GaAs buffer-layer with unintentionally doped $p$-type impurities or in the $p$-type Be-doped AlGaAs layer, or due to a slowly saturated magnetization component as is also seen in GaMnAs.

The $T_C$ of the heterostructure was estimated from the temperature ($T$) dependence of spontaneous magnetization $M(B=0)$ derived from the data of $R_H(B=0)/R_{sheet}$ for $T<T_C$, as shown by open circles in Fig.3(f). Also, the temperature dependence of inverse magnetic susceptibility $\chi^{-1}$ was derived from the data of $BR_{sheet}/R_H$ for $T > T_C$, where, the sample is paramagnetic, thus $M = \chi B / \mu_0$. Assuming that $R_O$ is $T$-independent in the Curie-Weiss equation[2], we plotted $\chi^{-1}$ as a function of $T$, as shown by solid squares in Fig. 3(f). By extrapolating the fit of $\chi^{-1} = \frac{T - T_C}{C}$ to cross the $T$-axis in Fig. 3(f), the $T_C$ value is estimated to be 192±2 K, which is in good agreement with the spontaneous $R_H/R_{sheet} - T$ data for $T < 200$ K (Fig. 3(a-e) and open circles in (f)). In this estimation, we set the Mn content $x = 0.25$ for $\theta_{Mn} = 0.5$ ML (since we defined the local Mn content as $\theta_{Mn}/2$ ML, where 2 ML is the width of Mn distribution along the growth direction estimated by TEM [2,11]). The Curie-Weiss plot fitted the experimental data assuming the skew-scatting as the dominant scattering mechanism ($R_{\text{S-sheet}} = cR_{\text{sheet}}$ [2]) better than assuming the side-jump scattering ($R_{\text{S-sheet}} = cR_{\text{sheet}}^2$).

We observed an increase of $T_C$ from our previous report of $T_C$=172 K in a similar I-HEMT-type heterostructure (0.3 ML Mn δ-doped GaAs/Be-doped $p$-Al$_{0.3}$Ga$_{0.7}$As [2]). The differences in structural parameters are the following: $\theta_{\text{Mn}}$ = 0.3 ML, Al content $x_{\text{Al}}$ in the $p$-AlGaAs is 0.3, and $d_s$ = 0 nm in the previous heterostructure, but in sample-A, $\theta_{\text{Mn}}$ = 0.5 ML, $x_{\text{Al}}$ = 0.5, and $d_s$ = 2 nm. Three important differences are that sample-A has a higher Mn concentration ($\theta_{\text{Mn}}$) in δ-doping, a higher GaAs/$p$-AlGaAs valence-band offset $\Delta E_V$ = 300 meV compared with 180 meV in the previous heterostructure due to the difference in $x_{\text{Al}}$, and an appropriate separation of $d_s$ = 2 nm. The sheet hole concentration $p$ is estimated to be 7.8±0.5×10$^{12}$ cm$^{-2}$ at room temperature from the ordinary Hall coefficient ($R_{\text{O-sheet}}$ = 80 Ω/T). This value is higher than the previous value $p$ = 2.2×10$^{12}$ cm$^{-2}$ [2]. The combined effects of increased Mn concentration, increased hole concentration, and enhanced 2DHG-Mn overlap led to the higher $T_C$ in the present I-HEMT structure of sample-A.

Next, we examined sample-B of Fig. 1(b), which is a N-HEMT type heterosturcture with Mn δ-doping. Figure 4(a) shows the temperature dependence of the sheet resistance $R_{\text{sheet}}$ of sample-B. The hump at around 250 K is attributed to the critical scattering nearby $T_C$[13], as was seen in (GaMn)As[14]. Figures 4(b)-(d) measured at 235, 240, and 250 K show clear ferromagnetic hysteresis in the Hall resistance ($R_H$–$B$) loops. This ferromagnetic hysteresis behavior changed to paramagnetic behavior at 260K (Fig. 4(e)) indicating $T_C \cong 250 \sim 255K$, which is in agreement with the hump in Fig. 4(a) at around 250K.

We ascribe the remarkably high $T_C$ of sample-B to both increased Mn concentration ($\theta_{Mn}$ = 0.6 ML) and reduced compensative Mn interstitials ($Mn_I$). It is pointed out that Mn atoms can occupy tetrahedral interstitial sites ($Mn_I$) in the ZB-structure and act as double donors[15,16]. The compensation of Mn acceptors at the Mn substitutional sites ($Mn_{Ga}$) by the $Mn_I$ donors decreases the hole concentration $p$ and $T_C$. The effect of the Fermi energy can have the significant contribution to the suppression of interstitial defect formation in the N-HEMT-type of sample-B, compared with that in the I-HEMT-type of sample-A. It is reported that the formation energies and thus the concentrations of charged (positive or negative) defects such as $Mn_I$ donors or $Mn_{Ga}$ acceptors are to a large extent controlled by the Fermi energy during growth[9]. In the I-HEMT structure in which the Mn δ-doped GaAs was grown after the growth of $p$-AlGaAs, $Mn_{Ga}$ formation becomes unfavorable during the Mn δ-doping growth because it further increases the hole concentration, and a high concentration of compensating $Mn_I$ defects are formed instead to balance the charge. This scenario turns opposite in the N-HEMT structure, making $Mn_{Ga}$ formation energetically favorable because the $p$-AlGaAs layer supplying holes was grown after the growth of the Mn δ-doped layer. Recently, Edmonds *et al.*[8] showed that the out-diffusion of the $Mn_I$ in (GaMn)As films strongly depends on the thickness of the film; with increasing the film thickness, longer LT-annealing at $\leq 200^oC$ is more effective to remove $Mn_I$ and to increase $T_C$. In fact, the sheet hole concentration of sample-B at room temperature was $p = 7.2\pm0.5\times10^{13}$ cm$^{-2}$[17], much higher than that of sample-A, leading to the metallic behavior of Fig. 4(a). It is also pointed out that the arsenic antisite ($As_{Ga}$) defects do not play a significant role in LT-annealing, since $As_{Ga}$ defects remain stable up to 450°C[18]. Therefore, coupled with the $Mn_I$ diffusion during the long LT-annealing treatment and the increase of Mn concentration, the above scenarios can explain the large difference of $T_C$ in the I-HEMT-type ($T_C$ = 192K) and N-HEMT-type ($T_C$ ~ 250K) heterostructures.

We observed a clear sign-change of AHE in sample B, as shown in Fig. 4(b-d). Fig. 4(f) shows the temperature dependence of the spontaneous Hall resistance $R_H$ (*B*=0T). Here, the spontaneous $R_H$ is defined as the Hall resistance at *B*=0T due to the spontaneous magnetization, after a positive magnetic field (*B*=0.5T) is applied and *B* is then reduced to 0T, as shown by the open circles in Fig. 4(b-e). As seen in Fig. 4(f), the spontaneous $R_H$ varies non-monotonously with *T*, and its sign is negative at $T \leq 240$ K, and positive at $T \geq 245$ K. The features of the sign-change and non-monotonous behavior in the $R_H$-*T* data could be due to the contribution of the Berry-phase effect to the AHE. Very recently, it has been suggested that the Berry phase effect in the crystal momentum space can also give rise to AHE[19-22] and can explain the AHE in (III,Mn)V magnetic semiconductors[19] and metallic SrRuO3[20]. Here, the Berry phase does not involve scattering, but depends on the Bloch states in the Fermi surface, and the Hall resistivity $\rho_H$ can be described as $\rho_H = -\rho^2 \sigma_H(M)$, where the Hall conductivity $\sigma_H$ does not depend on longitudinal resistivity $\rho$ and its dependence on *M* should be calculated from the band structure. The features of the sign-change and the non-monotonous behavior in the $R_H$-*T* data of sample-B in Fig. 4(f) are similar to those observed in metallic SrRuO3 ferromagnets, where first principal calculations of the Berry phase contribution to the AHE showed excellent agreement with the experimental data[20]. It is noteworthy that in other Mn δ-doped p-SDHSs (not shown here), we observed positive and monotonous behavior in the $R_H$-$T$ data in case of semiconductive samples as was the case of sample-A, and the sign-change and non-monotonous behavior become significant in metallic samples as was the case of sample-B. The difference between semiconductive samples and the metallic samples is in the nature of the quasi-particle's behavior; the metallic sample-B with higher hole concentration and higher Mn coverage probably has significant valence-band mixing with increased Fermi-surface curvature[19], leading to an observable Berry phase contribution to the AHE.

In the $R_H$-$B$ hysteresis loops of Fig. 4(b-d), step-like features appeared at around ±0.1 T. One might attribute such steps to the influence of domain-pinning in the MnAs clusters as the possible second phase. However, the possibility of MnAs cluster formation can be excluded by our study of HRTEM lattice images (for example, see Fig. 2(b) and (c)). Instead, the presence of an inclined magnetic easy-axis due to magneto-crystalline anisotropy of the system, as usually is seen in GaMnAs with in-plane magnetization easy-axis, could explain the origin of the step that appears during the magnetization reversal process under the external magnetic field perpendicular to the film. Another possibility is the influence of domain-wall-pinning at the patterned Hall bar edges during the motion of domain walls in magnetization reversal.

In conclusion, the large enhancement of $T_C$ using both I-HEMT and N-HEMT structures presented here can provide opportunities for fundamental studies in the magnetic quasi-two dimensional systems, and can also lead to functional device applications complying with the present semiconductor technology.

[17]  This value was estimated by assuming the same mobility $\mu = 1.9$ cm$^2$/Vs both in sample-A and sample-B, since the high $T_C$ of sample-B made it difficult to rule out the anomalous Hall effect contribution in Curie Weiss fitting to estimate $p$ and $\mu$. Our other Mn δ-doped *p*-SDHS samples (not shown here) of both I-HEMT and N-HEMT-type with relatively low $T_C$ are found to have similar $\mu$ values (2 ~ 5 cm$^2$/Vs), and thus supporting our assumption.

[18]  D. E. Bliss *et al.*, J. Appl. Phys. **71**, 1699 (1992).

[19]  T. Jungwirth, Q. Niu, and A. H. MacDonald, Phys. Rev. Lett. **88**, 207208 (2002).

[20]  Z. Fang *et al.*, Science **302**, 92 (2003).

[21]  Y. Yao *et al.*, Phys. Rev. Lett. **92**, 037204 (2004).

[22]  Byounghak Lee, T. Jungwirth, and A. H. MacDonald, Semicon. Sci. Tech. **17**, 393 (2002).

**Figure Captions**

**Fig. 1.** *p*-type selectively doped heterostructures (*p*-SDHS) grown by MBE on semi-insulating (SI) GaAs(001) substrates: **(a)** Mn δ-doped ($\theta_{Mn}$ = 0.5 ML) GaAs/Be-doped *p*-AlGaAs heterostructure (I-HEMT type) (sample-A). **(b)** Be-doped *p*-AlGaAs/Mn δ-doped ($\theta_{Mn}$ = 0.6 ML) GaAs heterostructure (N-HEMT type) (sample-B).

**Fig. 2. (a)** Cross-sectional HRTEM lattice image of a 1.0 ML Mn δ-doped sheet in GaAs grown under the same conditions of sample-A ($T_s$ = 300°C). The slightly dark area indicated by arrows corresponds to the Mn δ-doped sheet localized within a width of 2 – 3 ML. **(b-c)** Cross-sectional HRTEM lattice image of the Al$_{0.3}$Ga$_{0.7}$As/ GaAs heterostructure with $\theta_{Mn}$ = 0.6 ML (sample B: $T_s$ = 285°C). (c) is an enlarged image at the Mn δ-doped layer. There is no dislocation, no visible second phase or MnAs clusters, and the structure maintains the zinc-blende type crystal structure.

**Fig. 3. (a-e)** Hall resistance $R_H$ loops of the heterostructure of Fig. 1(a) at (a) 130 K, (b) 180 K, (c) 185 K, (d) 190 K, and (e) 205 K. **(f)** Temperature (*T*) dependent data of $R_H(B=0T)/R_{sheet}$ ($\propto M$) at $T<T_C$, and also of $BR_{sheet}/R_H$ ($\propto \chi^{-1}$) at $T > T_C$. The linear plot is $\chi^{-1}$ [=$(T-T_C)/C$], where $T_C$=192 K.

**Fig. 4**. **(a)** Temperature dependence of the sheet resistance $R_{sheet}$ of sample-B whose structure is shown in Fig. 1(b). **(b-e)** Hall resistance $R_H$ loops of sample-B measured at (b)

235 K, (c) 240 K, (d) 250 K and (e) 260 K. **(f)** Temperature dependence of the spontaneous Hall resistance $R_H$ at $B=0T$ (also shown by open circles in (b-e)).

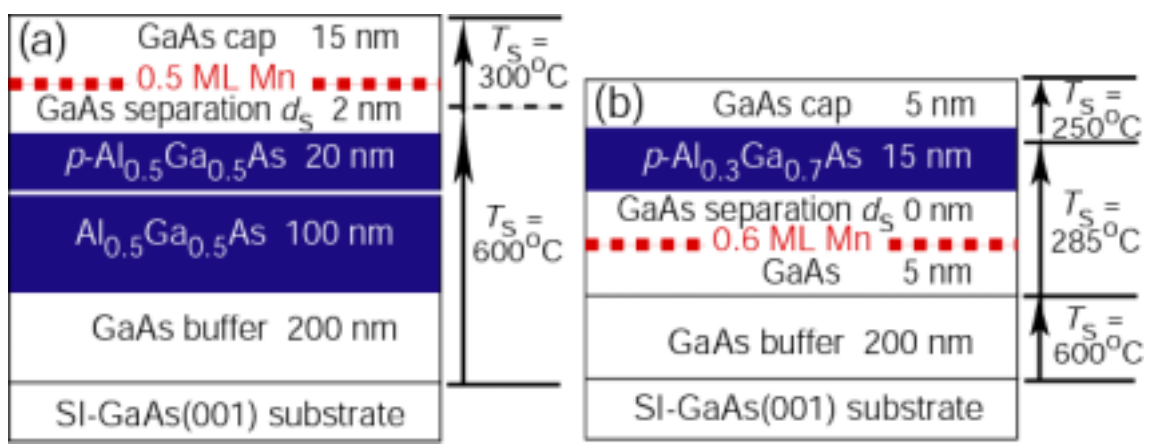

Fig. 1 Nazmul *et al*.

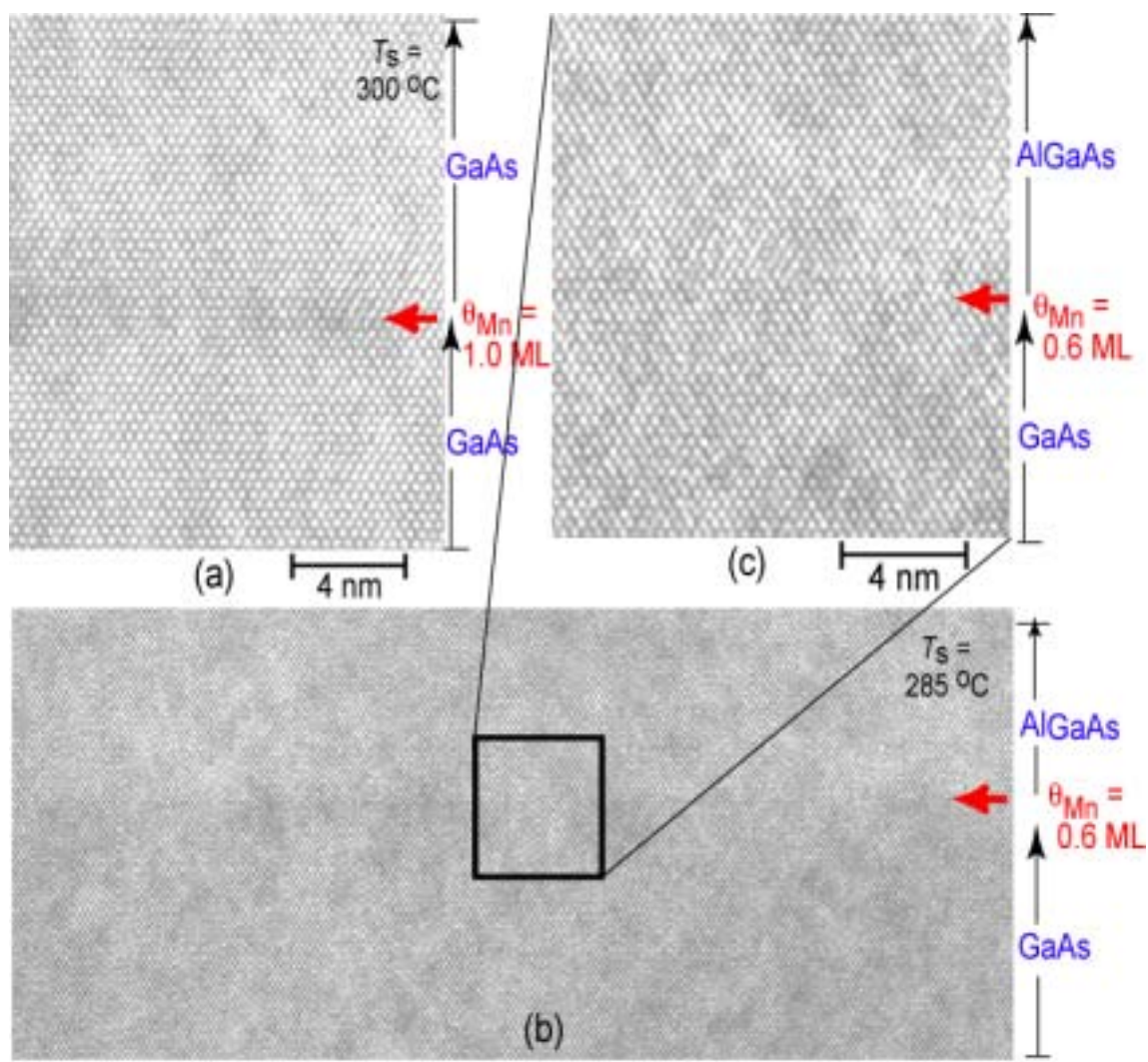

Fig. 2  Nazmul *et al*.

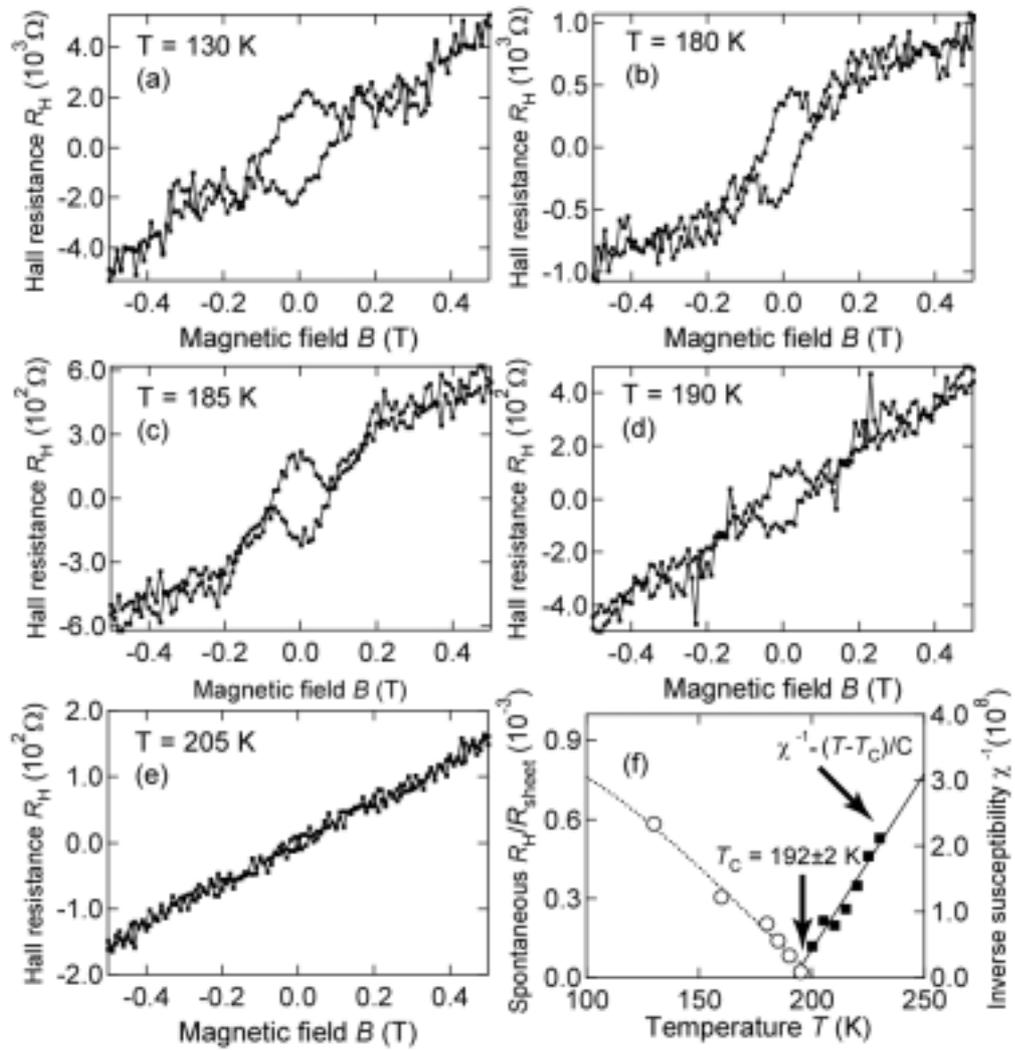

Fig. 3   Nazmul *et al*.

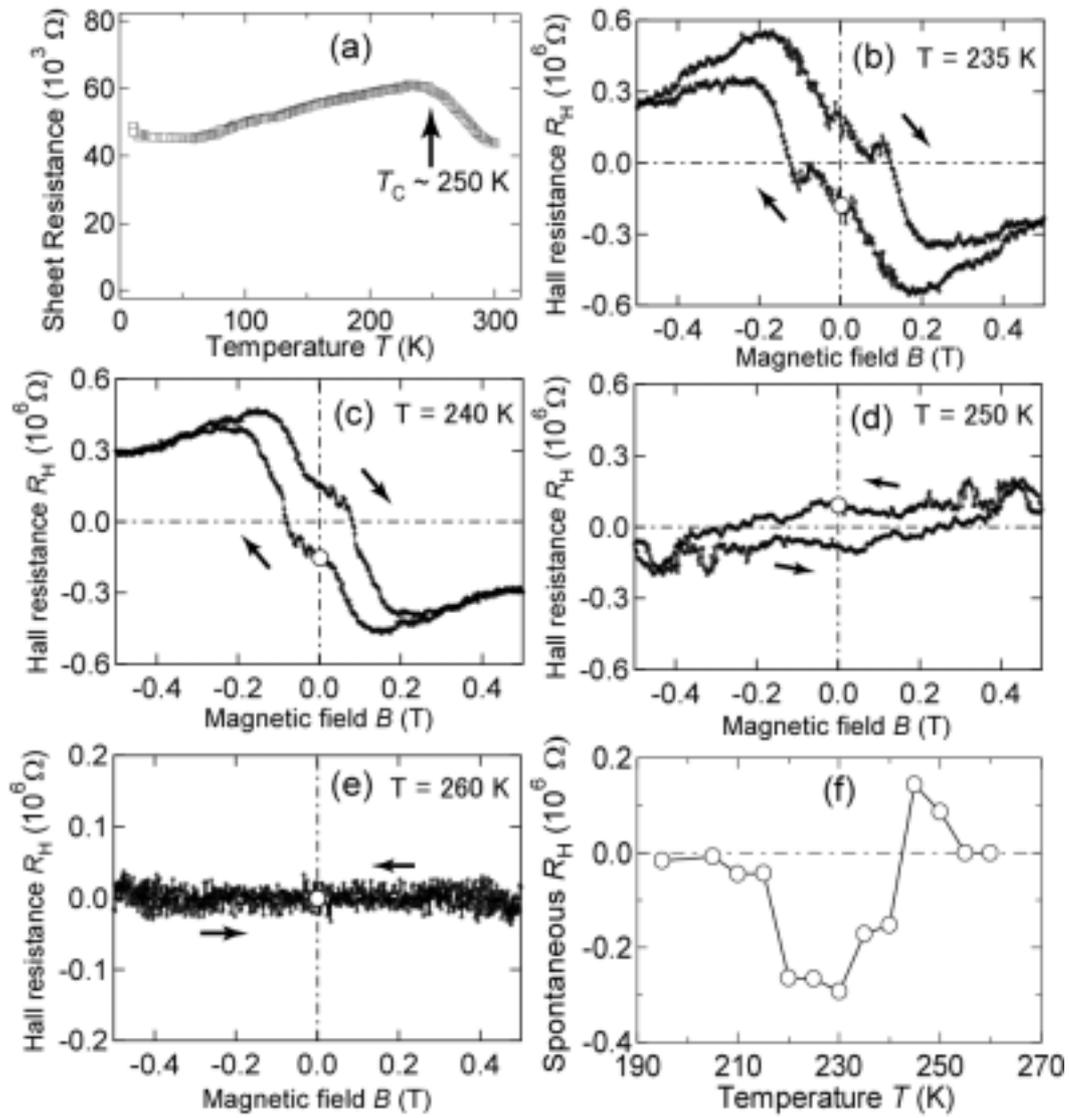

Fig. 4  Nazmul *et al*.

18